# Laser ablative fabrication of nanocrowns and nanojets on the Cu supported film surface using femtosecond laser pulses.

A.A. Kuchmizhak[1], D.V. Pavlov[1,2], Yu.N. Kulchin,[1,2] O.B. Vitrik[1,2]

[1]Institute of Automation and Control Processes, Far Eastern Branch, Russian Academy of Science, Vladivostok 690041, Russia
[2]Far Eastern Federal University, 8 Sukhanova str., Vladivostok 690041, Russia
*Corresponding author: ku4mijak@dvo.ru

Formation dynamics of the nanojets and nanocrowns induced on the surface of the Cu supported films of different thickness under the impact of tightly focused femtosecond pulses was studied in detail. We show that the single-shot fs-pulse irradiation of the 120-nm-thick Cu film results in formation of a single nanojet, which splits at increased pulse energy into two and then into a plurality of periodically arranged nanospikes eventually acquiring the form of the so-called nanocrown. The number of nanospike in the nanocrown was found to be linearly dependent on the pulse energy and nanocrown radius. The key role of subsurface boiling occurring on the metal film – substrate interface in the formation process of crown-like nanostructures was revealed by comparing the obtained results with the formation dynamics studied for thinner 60-nm and 20-nm-thick Cu films. In addition, the applicability of the fabricated nanostructures as low-cost substrate for photoluminescence signal enhancement of the organic dyes is also discussed in this paper.

*Keywords: laser nanostructuring, femtosecond pulses, nanojets, nanocrowns, hydrodynamic instability, subsurface boiling, photoluminescence enhancement.*

## 1. Introduction

Functional metal nanostructures demonstrating unique optical and spectroscopic properties are of great scientific interest in the past decade. Many promising applications of these nanostructures in such rapidly developing scientific and technological areas as plasmonics, nanophotonics, biosensors and superresolution microscopy were theoretically predicted and experimentally demonstrated [1-4]. Meanwhile, these areas are required both a single nanoelements accurately fabricated at a given point on the sample surface and relatively large ordered arrays including millions of nanoscale structures. Single metal nanostructure currently can be efficiently fabricated using advanced "top-down" approaches utilizing focused electron and ion beam pattering. However, these methods are relatively time-consuming and extremely expensive in terms of their applicability to fabrication of large ordered arrays of functional nanostructures, which initiates an active scientific search for alternative high-performing fabrication techniques. Ablative laser-assisted processing of metal films utilizing tightly focused nano- and femtosecond pulses is among promising solution of this task [5-12]. The impact of such laser pulses initiates a sequence of thermal [5-7], hydrodynamic [8,9] and electrodynamic [10-12] processes on the tens of nanosecond time scale. These often competing processes lead to formation of a large variety of functional nanostructures: resolidified nanojets [5,6,13], spherical nano- and mesoparticles [14,15], microbumps [16,17], micro- and nanoholes [18,19], nanocrowns [7,9], etc. It should be noted that the underlying physical mechanisms and processes describing the nanostructure formation dynamics and defining a particular type of fabricated nanostructure are the subject of intense scientific discussions and evidently require more detailed study. Therefore, new experimental data concerning the features of the nanostructures formation dynamics potentially can contribute to a comprehensive understanding of the physical processes and mechanisms, which occur with the metal films under single-pulses irradiation. This data can also points out the new ways towards creation of simple, cost-effective and high-performance laser-assisted techniques for fabrication of single functional nanostructures as well as their ordered arrays.

In this paper, we rigorously study the formation dynamics of the nanostructures (nanojets and nanocrowns) fabricated on the surface of the Cu supported films of different thickness under the impact of tightly focused femtosecond pulses. We show that the impact of fs-pulse on the 120-nm-thick Cu film results in the formation of a single nanojet, which splits at increased pulse energy into two and then into a plurality of periodically arranged nanospikes eventually acquiring the form of the so-called nanocrown. By comparing the obtained results with the formation dynamics studied for thinner 60-nm and 20-nm-thick Cu films we conclude the key role of subsurface boiling occurring on the metal film – substrate interface in the formation process of crown-like nanostructures. In addition, the applicability of the fabricated functional nanostructures as low-cost substrate for photoluminescence signal amplification of the organic dyes is also discussed in this paper.

## 2. Methods

In our experiments, laser nanostructuring (Fig. 1) of the Cu films was carried out using single linearly-polarized pulses generated at a wavelength of 532 nm through the parametric amplifier (TOPAS prime, Spectra Physics) pumped by Ti:Sapphire laser system (Spitfire Amplifier and Tsunami Femtosecond Oscillator, Spectra Physics). The output radiation from the parametric amplifier via the fiber coupler containing the micrometer positioner and the lens (NA = 0.25) was inputted into a segment of single-mode optical fiber (Thorlabs SM300) providing the radiation filtering into the spot with a nearly Gaussian intensity distribution profile, and then was focused on the sample surface by a high-NA lens (x100, NA = 0.95) under complete filling of their input aperture. This allows one to irradiate the samples by laser pulses with almost perfect lateral energy distribution profile at the focal spot with the size $R_{opt} = 1.22\lambda \cdot (2NA)^{-1} \sim 0.34$ µm.

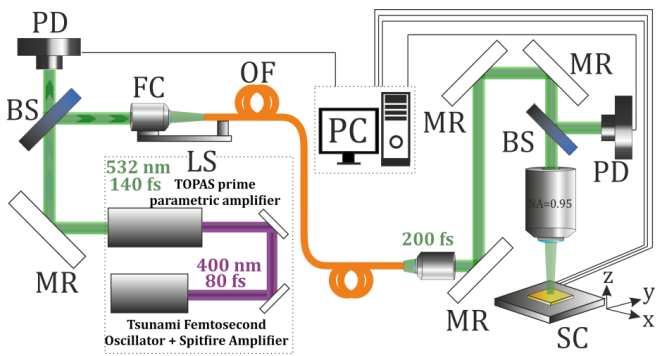

Fig.1. Schematic of the experimental setup for laser-assisted nanostructuring of Cu films: LS- laser system, MR – mirrors, BS – beam splitters, PD – photodetectors, FC – fiber coupler, OF – optical fiber, SC – scanning.

The sample was arranged on a PC-driven motorized micropositioning platform (Newport XM series) with a minimal translation step of 50 nm along three axes and moves from pulse to pulse. The pulse energy E was varied by means of a tunable filter and measured by a sensitive pyroelectric photodetector. Copper film with the thicknesses ~ 20, 60 and 120 nm deposited onto the optically smooth 2-mm-thick glass substrate by e-beam evaporation procedure (Ferrotec EV M-6) at a pressure of $5 \cdot 10^{-6}$ bar and an average speed ~8 Å/c were used as a samples for laser nanostructuring experiments. To increase the adhesion of the deposited material to the glass substrate, the latter was pre-cleaned by the build-in ion source (KRI EH200). The film thickness was measured at the preliminary step by calibrated piezoelectric resonator (Sycon STC-2002) mounted into the vacuum chamber. These measurements were controlled by the atomic force microscope (NanoDST, Pacific Nanotechnologies). The resulting Cu surface after fs-pulse irradiation was characterized by scanning electron microscope (SEM, Hitachi S3400N). All nanostructure in this work were fabricated using single-pulse irradiation under normal conditions.

Back-scattered dark-field optical images of the fabricated nanostructures were recorded using the high-resolution optical microscope (Hirox KH7700) equipped with a lens (NA = 0.8, x700) and a CCD-camera to record the optical images. The nanostructure illumination was performed by the broadband white light source coupled to the dark-field condenser minimizing the scattering signal from the flat Cu film surface. To measure the dark-field scattering spectra of nanostructures we used the optical microscope combined with a sensitive spectrometer (Shamrock 303i, Andor) equipped with a TE-cooled EMSSD-camera (Newton 971, Andor). The average back-scattering signal measured from the area containing a large number (up to 200) of similar nanostructures was normalized on the scattering spectrum measured from the smooth Cu film surface as well as on the white light source spectrum.

For surface-enhanced photoluminescence (PL) detection we used an organic dye of Rhodamine 6G (Rh6G) with an excitation peak near 532 nm. Alcohol solution of the R6G was coated the ordered array of the nanocrown by means of microinjector and after complete drying of the alcohol was obliquely (at an angle of 70° relative to the surface film normal) illuminated by the collimated beam from the semiconductor CW-laser source (Milles Griot) with a central wavelength of 532 nm. The pump radiation scattered from the nanocrown array and collected by the microscope lens was completely blocked by means of the long wavelength pass filter (with a band edge ~545 nm) placed in front of the microscope CCD-camera and provided the pump wavelength absorption of about $10^6$.

## 3. Nanocrowns and nanojets formation

SEM images illustrating the impact of the single femtosecond pulse onto the 120-nm-thick Cu film surface deposited on a glass substrate at increased pulse energy E are shown in Fig. 2a. As seen, at E = 4.6 nJ formation of a small microbump with a lateral size ~350 nm and the height ~200 nm is observed. Small increase of the pulse energy E up to 5.2 nJ substantially increases the lateral size of the modified area up to 600 nm accompanied by the accumulation of the molten film material at the microbump center and its subsequent resolidification in the form of so-called nanojet (Fig. 2a). Upon reaching the threshold energy E ≈ 5.5 nJ the microbump collapses forming the nanosized through-hole, while the nanojet splits into the two almost identical nanospikes. The diameter of the central through-hole increases with the pulse energy, wherein the molten copper film at its edges demonstrates the periodic modulation of the height resulting in the formation of multiple nanospikes. The nanospike number N was found to increase from 2 (at E ≈ 4.5 nJ) to 8 (at E ≈ 7.4 nJ) demonstrating nearly linear raise with the nanocrown radius $R_{crown}$ as well as with the pulse energy E (see. Fig. 2b). It is noteworthy that similar linear dependence of periodically arranged nanospikes N on the nanocrown radius was also observed in the case of ns-pulse irradiation of the Au film on a copper substrate [9] and fs-pulse irradiation of bulk Si surface [20], although the shape and the size of the obtained submicron spikes in the abovementioned cases was somewhat different. This presumably points out the universal nature of the hydrodynamic thermocapillary/thermocavitation (Marangoni, Rayleigh-Plateau) instabilities appearing in the molten material rim, despite the fact that formation processes of such rim for thin film and bulk samples obviously can significantly differ.

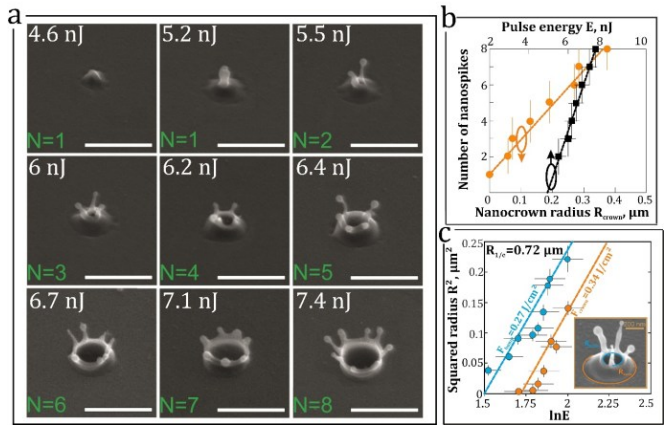

Fig.2. (a) Side-view (at an angle of 45°) SEM images illustrating the formation dynamics of the nanojets and nanocrowns on the 120-nm-thick Cu film under single-shot fs-pulse (~ 200 fs) irradiation at increased pulse energy E. N indicates the number of nanospikes in the nanocrown. The scale bar corresponds to 1 µm. (b) Dependencies of nanospikes number N on the pulse energy E and nanocrown radius $R_{crown}$. (c) Squared radius $R^2$ of the microbump and nanocrown versus natural logarithm of the pulse energy lnE (in nanojoules) measured for nanocrowns and microbump. The inset shows SEM image of the nanocrown with N=4 illustrating main geometrical dimensions $R_{bump}$ and $R_{crown}$.

The characteristic Gaussian radius of the surface energy distribution $R_{1/e}$ obtained via the linear slope of the dependence of the squared radii of the nanocrown and the microbump on the natural logarithm of the pulse energy lnE [21] was found to be approximately the same for both types of nanostructures and equal to $R_{1/e} = 0.72 ± 0.05$ µm (see. Fig. 2c). This value substantially exceeds both the Cu film thickness and the optical radius of the focal spot $R_{opt}$ ~ 0.34 µm, presumably indicating the lateral heat transfer on the nanosecond formation times of the microbump and the nanocrown $\tau_{crown}$. Indeed, based on the standard estimation of the thermal diffusion length $L_T = \sqrt{4\chi\tau_{crown}}$ at $\tau_{crown} \approx 10^{-9}$ [22] and the well-known high-temperature thermal diffusivity of the copper $\chi \approx 0.82$ cm$^2$s$^{-1}$ [23] (at T = 1300 K) one can obtain $L_T \approx 0.57$ µm. Then, the overall energy deposition radius was found to be equal to $R_T = (R_{opt}^2 + L_T^2)^{1/2} \approx 0.7$ µm [24], which is in good agreement with the experimentally measured $R_{1/e}$ value. The dependencies presented in Fig. 2c also indicate the threshold energy $E_{th}$ and the threshold fluence $F_{th}$ required for nanocrown formation. In accordance with the data presented these values are equal to $E_{th,crown} = 5.4$ nJ and $F_{th,crown} = E_{th}/\pi(R_{1/e})^2 = 0.34±0.04$ Jcm$^{-2}$, respectively. Similarly, for microbump one can estimate $F_{th,bump} = 0.27±0.03$ Jcm$^{-2}$ at $E_{th,bump} = 4.5$ nJ.

Taking into account the presented experimental data, one can conclude that the formation of the microbump and the nanojet (see. Fig. 2 (a)) at the first stage of the overall formation dynamics can be still described by well-established model [22,25], which includes the following sequence of physical mechanisms. First, laser pulse rapidly melts the Cu film. Second, the surface tension gradient (dσ/dT) caused by the lateral temperature gradient arising from the irradiation of the Cu film by the Gaussian-shaped laser beam accumulates the molten material at the center of the laser spot on the sample surface apparently coinciding with the temperature maximum. Third, the metal film detaches from the glass substrate, which is usually associated with thermal stresses in the metal film caused by fast molten film expansion, electron-lattice energy equilibration, etc. [22,25-27]. Finally, the molten material mainly concentrated at the microbump center resolidifies in the form of the nanojet (Fig. 2a). It should be noted that the sub-surface boiling process occurring at the "film-substrate" interface and initiating the intense vapor recoil pressure [28] represents an alternative mechanism describing the process of film detaching from the substrate.

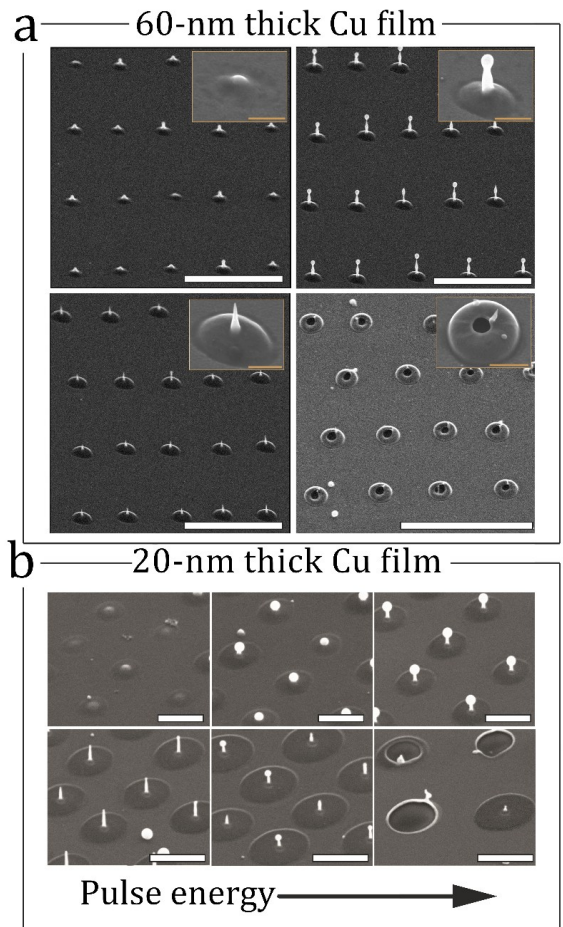

Fig. 3. SEM images of nanostructures fabricated in the 60- (a) and 20-nm-thick Cu films irradiated by single femtosecond (~ 200 fs) pulses at increased pulse energy E (E increases from left to right). Insets in (a) demonstrate enlarge SEM images of the typical nanostructures. The scale bars correspond to 5 and 1 microns for (a), (b) and insets, respectively.

Note also that the microbump destruction in the direction normal to the metal film surface accompanied by the nanojet splitting at the next stage of the formation dynamic presumably indicate significant increase of the factors causing the vertical movement of the molten material. In our opinion, this feature marks out the increased vapor recoil pressure in the vapor cavity at the "film-substrate" interface triggering the intense sub-surface boiling process. Microbump destruction is appeared to initiate the Marangoni hydrodynamic instability, which periodically modulates the molten rim increasing the number of nanospikes.

As it was discussed in [28], sub-surface boiling process in films with high thermal conductivity may be associated with a sub-surface temperature maximum, which appears on sub-nanosecond formation times owing to the balance of the thermal conductivity process and intensive evaporative cooling of the film surface. Under femtosecond pulse irradiation such maximum can be formed only in the relatively "thick" metal films (thicker than ~110 nm). In thin films, due to weak heat dissipation during ablation and sufficiently effective energy redistribution along the film thickness the sub-surface temperature maximum does not appear resulting in microbump collapsing without the nanocrown formation [28]. The latter argument is confirmed in our experiments on single-shot irradiation of "thin" Cu films with thicknesses of 60 nm (Fig. 3a) and 20 nm (Fig. 3b) under the same experimental conditions. As seen, melt dynamics for such films includes the formation and subsequent destruction of microbump accompanied by the spherical droplet ejection. For such relatively "thin" and chemically "pure" Cu films the formation processes exhibits the standard well-established behavior described earlier by numerous papers (mainly for thin Au films) [5,16,22,29] and suggested the absence of the sub-surface boiling mechanism. It also should be noted that subsurface boiling can appear even in "thin" 50-nm-thick metal films in the case of chemical inhomogeneity of the latter, as it was shown in [30] for Au/Pd alloy films demonstrating the appearance of nanojets and nanoparticles with porous internal structure during laser ablation process.

## Photoluminescence enhancement on the nanocrown array

Crown-like nanostructures demonstrated in this paper represent a set of nanospikes or nanojets, while the pulse energy E was found to be an effective and convenient parameter to control the number N of such nanosized protrusions. Recording density of such nanotips in the nanocrown array can be N times higher being compared with single nanojet array. For recorded array of the nanocrowns with the distance between the adjacent nanocrown ~2 μm (see Fig. 4a) the estimated recording density reaches ~ 6·10$^{12}$ nanospikes per square meter. Simple estimations show that the maximum nanospike density for a given material can be as high as ~10$^{14}$ nanospikes per square meter at the distance between the adjacent nanocrown ~0.7 μm. This feature opens up new perspectives for Cu nanocrown utilization as low-cost material in advanced vacuum devices for electron emission amplification [31,32].

Each nanospike in the nanocrown has the characteristic size comparable to the wavelength of visible light and consequently can act as an optical antenna concentrating the incident far-field radiation into the localized near-fields due to lighting rod effect as well as via the surface plasmon excitation. The dark-field back-scattering optical image of the nanocrown array (Fig. 4a) recorded under white-light illumination clearly show a bright yellow-red color presumably indicating the resonant scattering of the incident electromagnetic radiation on the nanocrowns. Average dark-field back-scattering spectrum measured from the array containing approximately 200 nanocrowns demonstrates a resonance peak (Fig. 4b) with central wavelength ~630 nm and the FWHM ~160 nm, which correlates well with the yellow-red scattering color appearing in the dark-field images (Fig. 4a).

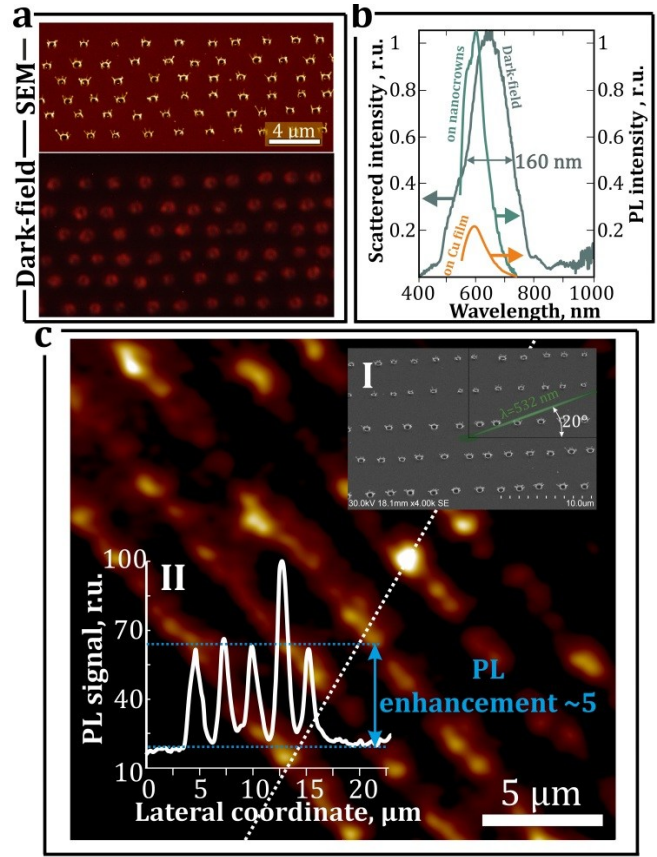

Fig. 4. Optical and spectral properties of Cu nanocrowns. (a) False-color SEM and back-scattered dark-field optical images of the nanocrown array. (b) Average dark-field back-scattering spectrum measured from the array containing approximately 200 nanocrowns and Rh6G PL spectra measured on the 120-nm-thick Cu film and on the Cu nanocrowns array. (c) PL images of the Rh6G deposited on the nanocrown array. Inset I shows SEM image of the corresponding area. Inset II shows PL intensity distribution measured along the direction marked by white dotted line. All SEM images were obtained at an angle of 45°.

A rather large FWHM value of the scattering peak presumably indicates the excitation of multiple plasmon resonances, the number of which due to the complexity of the nanocrown geometric shape is difficult to estimate using well-known analytical and numerical approaches [33]. Nevertheless, large FWHM value of the scattering maximum covering a part of the visible and near-infrared spectral ranges allows one to use nanocrowns as universal substrate for moderate PL signal enhancement for a variety of material including quantum dots, semiconductor nanocrystals, organic dyes, etc. Figure 4c demonstrates the PL image of the R6G layer deposited on the Cu film surface containing five rows of nanocrowns (Inset I in Fig.4c). As expected, the enhancement of the PL signal is observed near the nanocrown rows, which apparently indicates the increased electric-field intensity of the incident radiation due to the plasmon mode excitation. By comparing the PL intensity near the nanocrowns with the signal measured from the unmodified Cu surface (see Inset II in Fig.4c) as well as corresponding PL spectra one can find at least five-fold enhancement of the PL signal. Obviously, larger enhancement can be achieved by using organic dyes with excitation peak near the nanocrown scattering maximum (see. Fig. 4b).

## Conclusions

In conclusion, in this paper the formation dynamics of the nanojets and nanocrowns induced on the surface of the Cu supported films of different thickness under the impact of tightly focused femtosecond pulses was studied in detail. We show that the single-shot fs-pulse irradiation of the 120-nm-thick Cu film results in formation of a single nanojet, which splits at increased pulse energy into two and then into a plurality of periodically arranged nanospikes eventually acquiring the form of the so-called nanocrown. The number of nanospike in the nanocrown was found to be linearly dependent on the pulse energy and nanocrown radius. The key role of subsurface boiling occurring on the metal film – substrate interface in the formation process of crown-like nanostructures was revealed by comparing the obtained results with the formation dynamics studied for thinner 60-nm and 20-nm-thick Cu films. In addition, the applicability of the fabricated functional nanostructures as low-cost substrate for photoluminescence signal amplification of the organic dyes is also discussed in this paper.

## Acknowledgements

Authors are grateful for partial support to the Russian Foundation for Basic Research (Projects nos. 14-02-31323-mol_a, 14-02-00205-a, 15-02-50026-a) and the project of the RAS Presidium Program. The project was also financially supported by the Russia Federation Ministry of Science and Education, Contract № 02.G25.31.0116 of 14.08.2014 between Open Joint Stock Company "Ship Repair Center "Dalzavod" and RF Ministry of Science and Education. A.A. Kuchmizhak is acknowledging for partial support from RF Ministry of Science and Education (Contract No. MK-3056.2015.2) through the Grant of RF President.